\documentclass[a4paper,journal]{IEEEtran}  

\usepackage[english]{babel}
\usepackage{subcaption}
\usepackage[a4paper,left=1.57cm,right=1.57cm,top=0.95cm,bottom=2.54cm]{geometry}

\usepackage{amsmath}
\usepackage{amssymb}
\usepackage{braket}
\usepackage{graphicx}
\usepackage{hyperref}

\title{Tolerating Device Failure in Distributed Quantum Computing}

\author{\IEEEauthorblockN{Evan Sutcliffe\IEEEauthorrefmark{1} and Coral M. Westoby\IEEEauthorrefmark{1}} \\
\IEEEauthorblockA{\IEEEauthorrefmark{1}\textit{Nu Quantum Ltd.}, Cambridge, United Kingdom \\
\{firstname.lastname\}@nu-quantum.com}
}

\begin{document}
\maketitle

\begin{abstract}
It is desirable that a distributed quantum computer can operate despite the replacement or failure of its constituent components, allowing the reliability of the distributed system to exceed that of its subcomponents. We first show that when quantum error correction is performed over a modular quantum network, quantum devices can be swapped out or replaced, during operation, with minimal impact on logical error rates. We also investigate the ability of the toric and hyperbolic Floquet quantum error correcting codes to protect logical information under low rates of modular node failure. In particular, we show that under the proposed distributed quantum error scheme, the selected codes are able to maintain good logical error suppression during the failure of entire nodes. For catastrophic node failure of probability $p/100$, we suggest that a distributed toric code would outperform one implemented on a monolithic device below a physical error rate of $0.05\%$.
\end{abstract}

\section{Introduction}
Quantum computers are expected to provide up to an exponential speed up compared to classical algorithms, with relevance to areas such as quantum chemistry and prime factorisation \cite{Montanaro2016,Shors1994}. 
Solving such problems will require quantum computers that can operate fault-tolerantly. This means that quantum computers continue to function despite qubit errors caused by undesired environmental interactions or imperfect control.

To overcome these sources of error, quantum error correction (QEC) can be employed. When using QEC, \textit{logical qubits} are encoded in the state of many \textit{physical qubits}, meaning certain physical errors can be detected and corrected, without causing logical errors \cite{Roffe2019QECIntro}. Quantum codes can be described by the parameters $[[n,k,d]]$, where $k$ logical qubits are encoded in $n$ data qubits ($k<n$). The code distance $d$ describes the minimum weight of an operator that takes one code word to another, and represents the minimal number of physical errors required to cause a logical error \cite{Roffe2019QECIntro}. As the operations required to perform QEC are also error prone, QEC generally requires that operations are performed with a physical error rate below a \textit{threshold} value. Below this threshold, an arbitrarily low logical error rate can be achieved by increasing the distance of the QEC code \cite{fowler2012surface}, allowing for large and complex quantum algorithms to be executed reliably.

\begin{figure}[h!tbp]
    \centering
    \includegraphics[width=\linewidth]{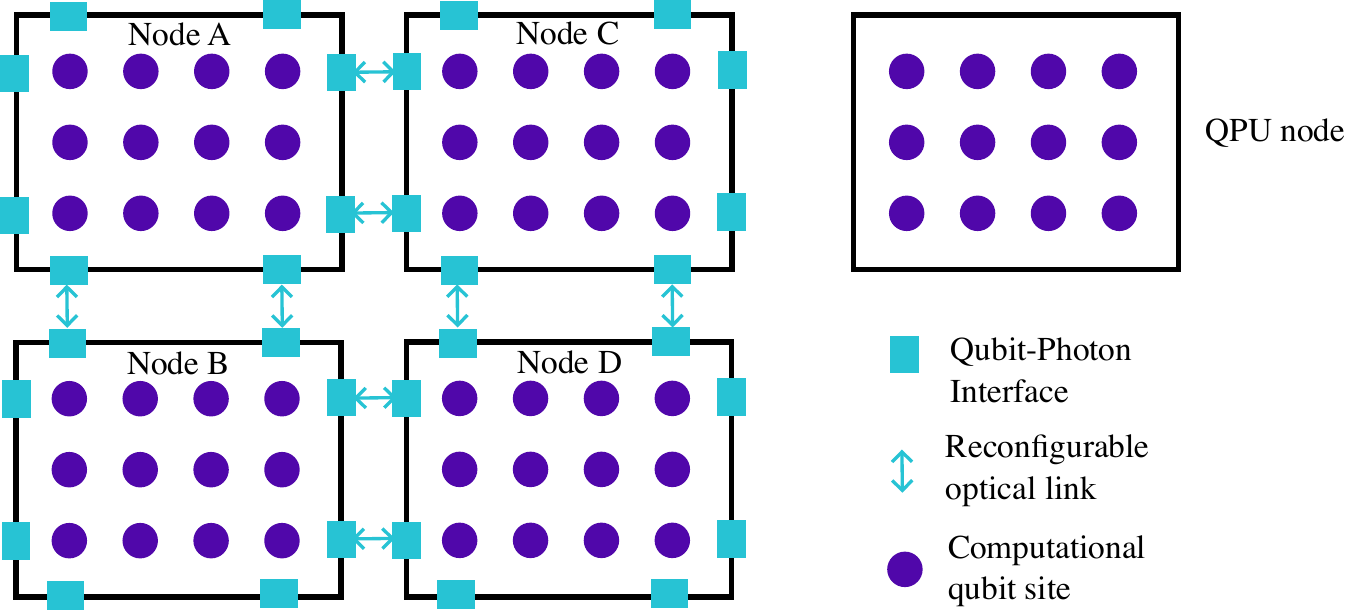}
    \caption{A quantum network to implement a distributed QEC code. A single distributed QEC code is then encoded over the qubits held by the set of QPUs. These devices are interconnected by QPIs, between which Bell states can be generated. The planar layout has been chosen for demonstration, and is not a requirement.}    \label{fig:placeholder_distq}
\end{figure}

Utility-scale fault-tolerant computers are expected to require between $10^5$-$10^6$ physical qubits due to the large overheads in physical qubits required to implement a quantum code \cite{Blunt2022Perspective,gidney2025factor,webster2026pinnacle}. Therefore, a second challenge for a fault-tolerant device is scaling up the number of physical qubits while maintaining physical error rates that are below the error threshold. 
Scaling up the number of qubits within a single device (termed \textit{monolithic scaling}) is regarded as a technically challenging hardware problem, requiring development of new techniques to resolve engineering challenges arising for every step up in scale \cite{saffman_quantum_2016,sarovar_detecting_2020,ang2024Multinode}. One potential solution is to consider \textit{distributed} QEC, in which the size of computation can be scaled by interconnecting smaller, modular devices, within a quantum network. In such an approach, the number of physical qubits can be scaled up by adding additional devices to the quantum network. An example quantum network constructed of Quantum Processing Units (QPUs) as network nodes is shown in Fig.~\ref{fig:placeholder_distq}. Here, each QPU is a device each holding an intermediate number of physical qubits. For distributed quantum error correction, the number of physical qubits can be increased by simply adding additional QPUs to the network. The QPUs are interconnected though quantum channels by Qubit-Photon Interfaces (QPIs) \cite{SussexQPI, InnsbruckQPI,sutcliffe2025distributed}. These QPIs represent additional qubits solely for communication, which can be used for generating non-local Bell states between separate devices over the quantum network.

Distributed QEC has been considered for a range of QEC codes, including surface codes \cite{nickerson2013topological,deBone2024thresholds}, colour codes \cite{chandra2025distributed}, and Floquet codes \cite{sutcliffe2025distributed}. To maintain a QEC code over a network, shared entangled states can be used as a resource for quantum communication \cite{Caleffi2022}.
One manner in which distributed QEC can be performed is by using a shared Bell state (entangled two-qubit state) as a resource for gate teleportation \cite{ Ramette2024,chandra2025distributed}. In such an approach, a standard syndrome extraction circuit can be utilised, but with operations performed between qubits held at separate devices implemented through the use of shared Bell resource states as well as local operations and classical communication \cite{Bennett1996}.

Another approach is to use an entangled state directly for performing syndrome extraction. For general quantum codes, a check of weight $w>2$, performed between data qubits held over $N\geq2$ separate devices, can be performed using a shared $N$-qubit GHZ state (${|\text{GHZ}\rangle}_N = \frac{1}{\sqrt{2}} (\ket{0}^{\otimes N}+\ket{1}^{\otimes N})$ \cite{deBone2020}. Here, each qubit of the shared GHZ state can be used in place of an ancillary qubit in the $\ket{+}$ state to perform syndrome extraction \cite{nickerson2013topological,deBone2024thresholds,singh2025modular}.\footnote{In a monolithic system, using a GHZ state for syndrome extraction is referred to as Shor-style extraction \cite{shor1996fault}.} When the qubits of the GHZ state are distributed across multiple devices, this allows for the syndrome extraction circuit to be performed using only the shared GHZ state. It has been shown that a distributed surface code can be implemented using small, modular devices ($O(5)$ qubits) that operate to generate shared 4-qubit GHZ states to perform weight-four syndrome measurements \cite{nickerson2013topological,deBone2024thresholds}. In a quantum network, GHZ states can be generated by combining a set of $N-1$ Bell states in a process known as entanglement fusion \cite{deBone2020}. Additionally, entanglement distillation can also be performed to improve the average GHZ state fidelity, at the expense of additional Bell state generation overheads \cite{nickerson2013topological,deBone2024thresholds}. 

One particular family of QEC codes that has been shown to perform well in a distributed setting is Floquet codes\cite{sutcliffe2025distributed}. These are a family of QEC codes in which the logical qubits are encoded dynamically. While they do not contain a static set of stabilisers, by performing \textit{check operations} in a known periodic schedule, logical qubits can be preserved and operated on \cite{hastings2021dynamically,moylett2025logical}. 

In Floquet codes, the syndromes are inferred from the product of weight-two checks, meaning that Floquet codes can perform well using qubit hardware that allows for native two-qubit parity measurements \cite{Gidney2021honeycomb, Haah2022BoundaryHoneycomb}. Floquet codes, therefore, have particularly efficient syndrome extraction operations in a distributed approach, where a single shared Bell state can be used to perform a weight-two check. Further, QEC codes with non-planar connectivity can be implemented in a distributed QEC approach by utilising the arbitrary long-range connectivity of entanglement generation in the quantum network, which has no planarity constraints. As such, high-rate codes such as hyperbolic Floquet codes can be implemented, using modular devices that themselves only have planar qubit connectivity \cite{higgott2024constructions,sutcliffe2025distributed}. 

QEC codes are commonly benchmarked by their ability to correct errors that occur on physical qubits with small, independent probabilities. However, errors that affect a large number of qubits simultaneously are also expected to be an issue for early quantum devices. For example, catastrophic errors might entirely erase a subset of qubits contained within a module \cite{xu2022distributed,coble2025correctionchainlossestrapped}.  One example of a catastrophic error is ion chain loss, a non-Pauli error process seen in trapped ion systems. Catastrophic errors can also be seen in superconducting systems, where non-Markovian errors can be caused by quasi-particle poisoning from cosmic rays. Correlated, high-weight, or non-Markovian errors are outside of the standard error model assumed when designing a QEC code, and can reduce the effective error distance. Node failure rates will strongly depend on the modality and engineering trade-offs chosen for any specific instantiation of a distributed quantum computer, but over a long enough timescale, the use of subsystem failure tolerance has been shown for classical communication and computational systems to be a vital part of constructing useful systems \cite{pease_reaching_1980,ansel_dmtcp_2009,armstrong_making_2003}. In a distributed system, it is highly desirable for the mean time between failure for the whole system to exceed that of any specific component. 

In distributed quantum computing, logical information is encoded over multiple separate devices, which raises the possibility that device failure events could be tolerated.
In a distributed setting, whole node failure describes a correlated error that occurs simultaneously on all qubits held on a single device, including non-Pauli errors due to events such as high energy cosmic rays or qubit loss. Node failure has been considered in the context of distributed Bivariate Bicycle codes, \cite{coble2025correctionchainlossestrapped}, and whole-device surface code patches, interconnected in a quantum network \cite{xu2022distributed}. In Coble \textit{et al.}, chain losses are shown to be tolerated, given that they occur with a low probability and on a small subset of qubits used to encode the Bivariate Bicycle code. However, only single chain loss events were considered, with tolerance to simultaneous loss events not considered. Another framework was considered by Xu \textit{et al.}, where each QPU is assumed to hold a separate surface code patch. These separate logical qubits are used as the base code for concatenation into an erasure-correcting QEC code. This approach requires each QPU to be sufficient to operate fault-tolerantly. However, such concatenation approaches introduce large qubit overheads to implement. Correlated errors have also been considered more generally, in the context of time-correlated errors \cite{f2025detrimental} and spatially-correlated errors \cite{rojasarias2026scalingsiliconspinqubits}. These show that in the worst-case, where correlated errors can affect all qubits in a QEC code, correlated errors can prevent the desired exponential suppression of logical errors.

Another major issue that can limit the performance of quantum devices is scheduled downtime, such as would be required for regular calibration or component replacement. For certain qubit modalities, these calibration cycles are currently performed with up to an hourly frequency, limiting the depth of executable quantum algorithms \cite{shnaiderov2025optimal,ibm_calibration}.

\begin{figure*}[tbp]
  \centering
  \begin{subfigure}[b]{0.45\linewidth}
  \begin{center}
    \includegraphics[scale=0.4]{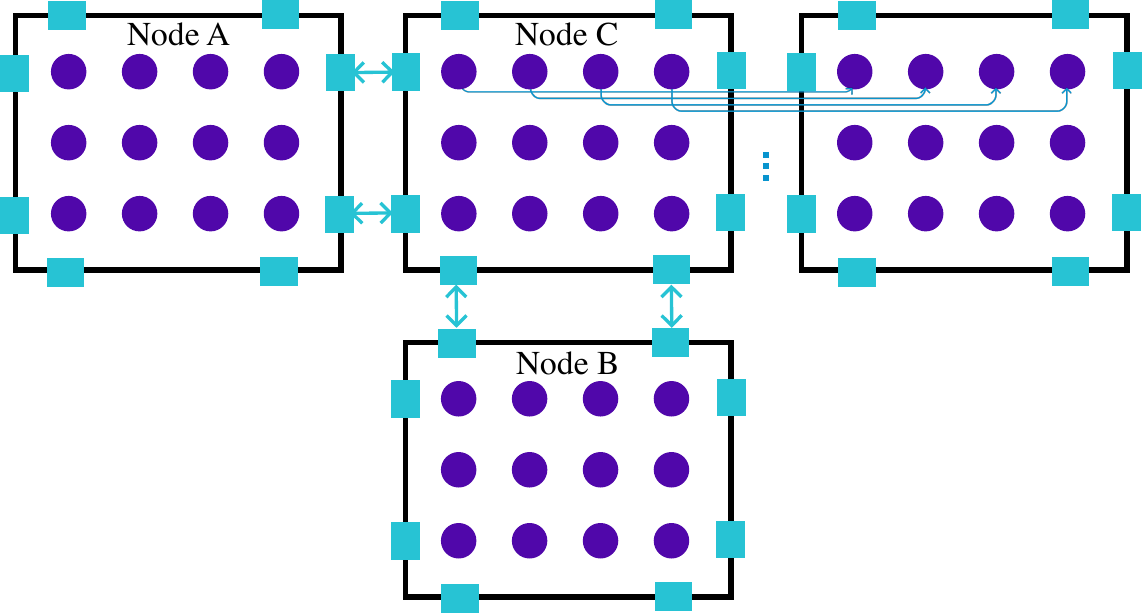}
   \end{center}

    \caption{}
    \label{fig:teleport_a}
  \end{subfigure}
  \hfill
  \begin{subfigure}[b]{0.45\linewidth}
    \begin{center}

    \includegraphics[scale=0.4]{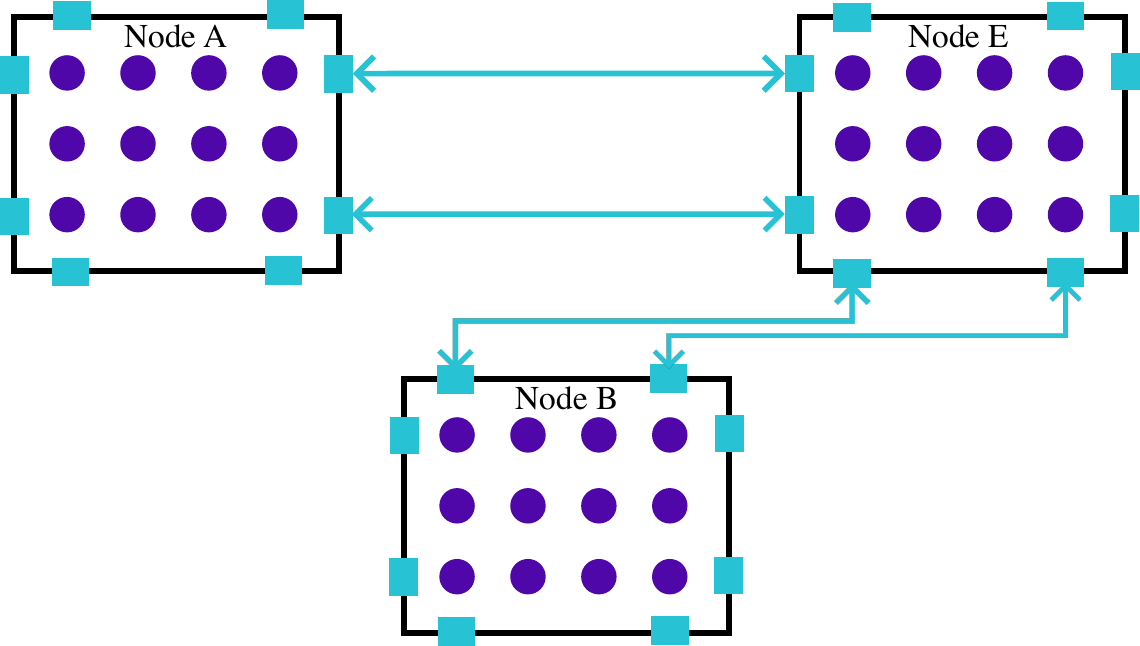}
    \end{center}

    \caption{}
    \label{fig:teleport_b}
  \end{subfigure}
  \caption{Schematic showing how a node (e.g. a QPU) can be swapped out of a quantum network while maintaining a QEC code. (a) For swap-out, data qubits held on Node \textbf{C} are each teleported out onto the unused node, Node \textbf{E}, using shared Bell states. Then, as shown in (b), the network is reconfigured such that Node~\textbf{E} has required connectivity to replace Node~\textbf{C} in the distributed QEC code.}
  \label{fig:swapout}
\end{figure*}
We generalise previous analysis on node failure in QEC to consider two types of node dropout: scheduled node downtime and unscheduled node failure. The term dropout is used to refer to either downtime or loss events, while we use node failure to refer to only unscheduled node loss.
For scheduled node downtime, we show how quantum information can be moved ahead of time using a qubit teleportation scheme \cite{bennett1993teleport}. For two different QEC codes, we further show that such an approach has a minimal impact on logical error rates.
The two code families considered are a distributed implementation of the toric code \cite{Kitaev2003ToricCode} and distributed hyperbolic Floquet codes \cite{higgott2024constructions,fahimniya2025fault,sutcliffe2025distributed}.
We also show that the proposed distributed QEC implementation allows for a tolerance to unscheduled node failure with minimal additional qubit overheads, and without modification to the decoding or syndrome extraction processes. This is in contrast to a monolithic system, in which node failure always results in the loss of quantum information. The results are shown to generalise to different QEC codes, as long as the number of qubits on any single device is small in comparison to the total size of the code.

\section{Methods}

In a distributed QEC code, qubits of the code are distributed among a set of QPUs, each of which has only $n_q$ ($n_q < n$) physical qubits. Syndrome extraction circuits that are performed between data qubits on two separate QPUs can be implemented using shared Bell states. For QEC codes with syndromes of higher weight checks, such as the toric code, distributed syndrome extraction circuits are performed using a shared $N$-qubit GHZ state. That is, stabilisers are measured between qubits held at $N$ separate devices using a GHZ state in place of the ancillary qubit (where each QPU holds a single qubit of the GHZ state). 

We consider QEC codes which are distributed over many individual quantum computers (referred to as QPUs), as shown in Fig.~\ref{fig:placeholder_distq}. Each QPU contains a small number of qubits ($16\leq n_q \leq 48$) and is equipped with $\lceil\sqrt{n_q}\rceil$ QPIs. We assume these QPIs are interconnected with a fully connected photonic routing network \cite{sutcliffe2025distributed}, allowing entanglement to be generated between arbitrary QPUs in the network.

Each QEC code is partitioned onto a distributed system by assigning each data and ancillary qubit to a vertex of a graph, with edges that describe the required qubit connectivity. In static codes, this graph is the Tanner graph \cite{Roffe2019QECIntro}. The graph is partitioned using spectral partitioning to give a set of clusters that all meet a maximum size constraint \cite{mcsherry2001spectral,sutcliffe2025distributed}.
Each cluster is assigned to a QPU, and check operations that span between separate QPUs are implemented using shared entanglement.

We consider a distributed implementation of hyperbolic Floquet codes, as described in \cite{sutcliffe2025distributed}. For example, the \hbox{H144-$f2$} code ($[[576,20,12]]$) is a Floquet code defined on a genus-10 semi-hyperbolic surface, allowing for $20$ logical qubits to be encoded. The base code is constructed using $144$ data qubits, has an encoding rate of $k/n \approx 1/8 $ logical qubits per physical qubit, and a code distance $d=6$. The parameter $f=2$ denotes that the code has been finegrained to finegraining level $f=2$ \cite{higgott2024constructions,sutcliffe2025distributed}, and hence is semi-hyperbolic. After finegraining, which is a process to increase the code distance as defined in \cite{higgott2024constructions}, the \hbox{H144-$f2$} code requires $144 \times f^2 =576$ data qubits and has distance $d=12$.\footnote{The finegraining process has no impact on the number of logical qubits $k$ encoded.} Also considered were distributed implementations of the unrotated toric code \cite{Kitaev2003ToricCode}. For both codes, decoding is performed using minimum weight perfect matching \cite{higgott2022pymatching}. For the node dropout simulations, decoding was performed with correlated error channels excluded from the detector error model. Decoding using this approach was found to perform accurately, despite the presence of large correlated errors.

Numerical simulations are run with 32 noisy rounds of error correction ($r = 32 $). Four additional noiseless rounds are run, two at the start and two at the end of the simulation, to minimise the impact of temporal boundary effects, which we do not consider in this work.

\begin{figure*}[tbp]
  \centering
  \begin{subfigure}[b]{0.49\linewidth}
    \begin{center}
    \includegraphics[scale=0.4]{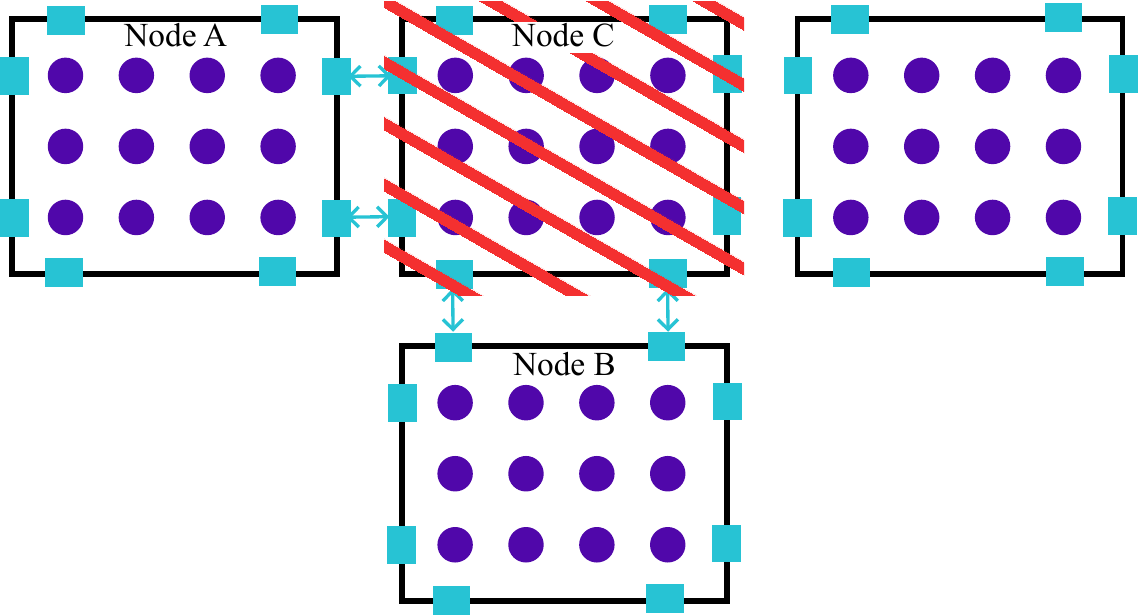}
    \end{center}
    \caption{}
    \label{fig:nodefail_a}
  \end{subfigure}
  \hfill
  \begin{subfigure}[b]{0.49\linewidth}
  \begin{center}
    \includegraphics[scale=0.4]{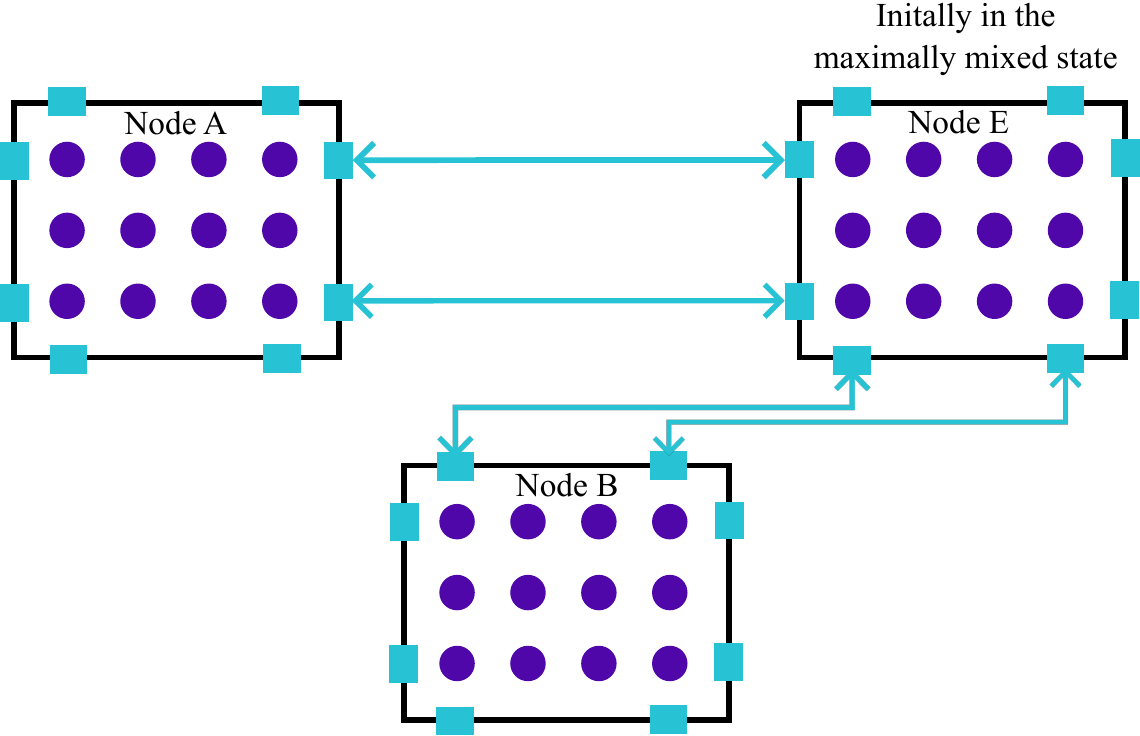}
  \end{center}
    \caption{}
    \label{fig:nodefail_b}
  \end{subfigure}
  \caption{The process of node swap-out when node failure is unscheduled. (a) A catastrophic failure occurring on Node~\textbf{C}. As such, the state of the qubits held at \textbf{C} are corrupted and discarded. Then, as shown in (b), the same number of (data) qubits at Node~\textbf{E} are initialised in the maximally mixed state, and Node~\textbf{E} is interconnected in place of \textbf{C}.}
  \label{fig:nodefail}
\end{figure*}

\subsection{Operational Model}
Firstly, we consider scheduled node dropout, where it is known ahead of time that a node is going to be taken offline, such as for calibration.

Figure~\ref{fig:swapout} shows the process of node swap-out, where the function of Node~\textbf{C}  in the network is replaced by a new, unused Node~\textbf{E}. Swap-out can be performed by using quantum teleportation to transfer the state of each data qubit held at \textbf{C} onto qubits held by \textbf{E}. Once this has been performed, the quantum network connections must also be reconfigured, such that \textbf{E} has the same connectivity as the node it is replacing. In a network of devices connected by optical channels, reconfiguration can be achieved with minimal overhead using optical switching \cite{sutcliffe2025distributed,shapourian2025quantum}. Performing qubit teleportation on the physical qubits in this manner is a transversal operation, meaning it is performed on each data qubit individually. Note that this approach differs from logical teleportation, as it is a subset of physical qubits that are being teleported, not a logical state. We consider node swap-out to be performed in a single step, between two QEC code rounds. However, if required, this approach could be performed over multiple code rounds with a subset of qubits teleported after each code round. This approach may introduce additional long range connectivity requirements, and was not considered in this work. 

For an unscheduled node failure, there is no opportunity to teleport data qubits into a replacement QPU ahead of time. Therefore, the information of all qubits held on that device is lost. We assume that node failure is detected within a single code cycle, either through sideband information or analysis of the error syndromes. Figure~\ref{fig:nodefail} shows the process of replacing a failed node with a functioning node. For this process, we assume the QEC code is distributed over many independent QPUs, such that the qubits held at other QPUs can still support the logical qubits. In this manner, node failure can be considered as simply a spatially correlated error. The Node~\textbf{E} is substituted into the network, in place of Node~\textbf{C}. The connections of \textbf{C} with the qubits held at \textbf{E} are initialised in the maximally mixed state $\mathbb{I}_d$, represented by the identity matrix with $d=2^{n}$. This represents the complete loss of quantum information previously encoded in these qubits. Syndrome extraction is then performed using qubits at \textbf{E} as substitutes to the corrupted qubits held at \textbf{C}. As the data qubits at \textbf{E} are initialised into a fully mixed state, syndrome extraction will initially show them to be erroneous. As long as the subset of qubits corrupted is small in comparison to the code distance, such approaches can be corrected using standard decoding processes. Smaller QPUs, however, will also require more non-local operations, resulting in additional entanglement overhead and more noisy qubit operations. This elevated noise rate will impact the ability of the quantum code to correct circuit-level noise, resulting in trade-offs in the optimal size of QPU.

\section{Failure and Noise Models}

\subsection{Noise Model} \label{sec:noisemodel}
Quantum error correction describes a process that can be used to correct physical errors due to undesired interactions with the environment or non-ideal qubit control. As such, the performance of such QEC codes will be highly dependent on the specifications of such errors. When modelling physical errors, we therefore use a fixed noise model to allow for accurate benchmarking. We consider a modified circuit-level noise model, in which all quantum circuit operations are non-ideal, and introduce physical errors \cite{Gidney2021honeycomb}. A common noise model used is Standard Depolarising 6 (SD6) described by Gidney \textit{et al.} \cite{Gidney2021honeycomb}. For this model, noisy circuit gates are simulated by applying the ideal gate followed by depolarising noise with probability $p$. We will henceforth refer to local operations (e.g., single-qubit operations or two-qubit gates between qubits on the same device) as `local' operations, with the noise described by a local error probability $p_l$. All local errors are assumed to be applied with uniform probability $p_l = p$. However, for distributed QEC, the noise model must also be modified to account for the elevated noise for performing operations between separate QPUs. Such `non-local' operations are mediated by shared Bell states, which should be expected to have a lower fidelity compared to operations performed between qubits within the same device. We made use of a distributed QEC noise model, described in \cite{sutcliffe2025distributed}. In this model, Bell states are modelled as being initialised ideally, followed by the application of two-qubit depolarising noise (termed non-local noise probability) with probability of $p_{nl} = 10p$ \cite{cambell2007bellpump}.\footnote{A Bell state modelled in this manner will have fidelity $\mathcal{F_\text{Bell}} = 1 - \frac{12}{15}p_{nl}$.}
Non-local operations are performed using a shared Bell state (for weight two checks) or a shared GHZ state (for checks between $N>2$ devices).
We assume all gates take unit time $\tau = 1$. Non-local Bell state generation, which can occur in parallel using $\lceil\sqrt{n_q}\rceil$ QPIs per node, takes time $\tau = 5$. Cumulative idling errors are applied at the end of each code round with a rate $p_l$ for each time-step for which a qubit was idle. For the toric code, measuring certain non-local stabilisers requires the generation of shared GHZ states. These are assumed to be generated from fusing a set of $N-1$ Bell states. The quantum circuits were constructed and simulated using the quantum error correction package Stim \cite{Gidney2021stim}.

\subsection{Node failure}
Node failure describes the catastrophic loss of all qubits held within a single device. This is modelled as the full depolarisation of all qubits held on that specific device. Node failure is assumed to occur with probability $p_{\text{dropout}}$ at the end of each code round. We assume node failure is a heralded event, where all quantum information has been lost or the node is no longer operational. Therefore, after a node failure event, a new node can be substituted into the network. Approaches which can herald a catastrophic failure have been described in Coble \textit{et al.} for chain losses \cite{coble2025correctionchainlossestrapped}. We assume the qubits of the new QPU are initialised in the maximally mixed state.

For a node containing $n_q$ qubits, node failure can be modelled using an ${n_q}$-qubit depolarising channel. That is, a Pauli operator from:
\begin{equation}
    \mathcal{P}_{n_q} = \{I,X,Y,Z\}^{\otimes n_q} / I^{\otimes n_q}
\end{equation}
is applied, with each of the $(4^{n_q}-1)$ possible error channel being sampled with uniform probability $p_{\text{dropout}}/(4^{n_q}-1)$ \cite{Gidney2021stim}.

However, as the number of possible Pauli errors grows exponentially with the $n_q$, it is impractical to exactly simulate correlated errors of this form for more than $\approx 10$ qubits (for a 10-qubit depolarising channel, there are $4^{10} \approx 10^6$ possible Pauli error channels to sample from). As such, we sample from a finite subset of possible Pauli errors from $\mathcal{P}_{n_q}$ to approximate the maximally mixed state. A set of $e$ Pauli errors are applied as a correlated noise channel, with each of the Pauli errors applied with effective probability $p_{\text{eff}} = p_{\text{dropout}}/e$. Using this approach, $e = 512$ was found to give accurate estimates for the depolarising noise channel.

When simulated in Stim, correlated errors are limited to being applied to at most 64 qubits \cite{Gidney2021stim}. Therefore a modified approach from that described in Coble \textit{et al.}, was used to model node failure in larger, e.g. monolithic systems, \cite{coble2025correctionchainlossestrapped}. This approach considers an ensemble of quantum circuits, with fully depolarising noise representing node failure applied at fixed locations in the quantum circuits. This is modelled by applying fully depolarising single-qubit noise to the $n_q$ qubits of a QPU. An ensemble of such circuits are then used to estimate the logical error rate under random node failure. This approach was only used to estimate the logical error rates for the monolithic system described in Fig.~\ref{fig:new_plot} in Section~\ref{sec:drop_out2}, as it was found to be more computationally intensive, with approximately 8000 separate circuit simulations required for the given confidence ranges. Confidence intervals were estimated using a bootstrapping process \cite{diciccio1996bootstrap}. This differs from the standard estimation of confidence used, as implemented by sinter \cite{Gidney2021stim}, but both techniques were used to calculate a 99.9\% confidence region.

In this work we examine $p_{\text{dropout}} \in [10^{-6},10^{-4}]$. Assuming a $10$ms cycle time, as might be anticipated on an ion trap platform, a $10^{-6}$ failure rate per check round implies a several hour mean time between failure at the node level, suggestive of recalibration. A failure rate of $10^{-4}$, suggests a mean time between failure on the scale of minutes which may be expected in systems subject to ion chain loss.

Node failure events are modelled as occurring per code round, unlike qubit errors, which are applied probabilistically after each qubit gate. However, this approach can be considered as being (phenomenologically) equivalent to modelling node failure as occurring after each physical gate cycle in a code round, with a different probability. This equivalence is because a node failure will randomise all readout measurements, local and distributed, if it occurs before readout. All QEC code circuits used in this work are assumed to perform all readout measurements simultaneously, at the end of each check cycle.  

\section{Results}

\subsection{Performance with scheduled node swap-out} \label{sec:swap_out}

\begin{figure}[h!tbp]
  \centering
  \begin{subfigure}[b]{0.95\linewidth}
    \includegraphics[width=\textwidth]{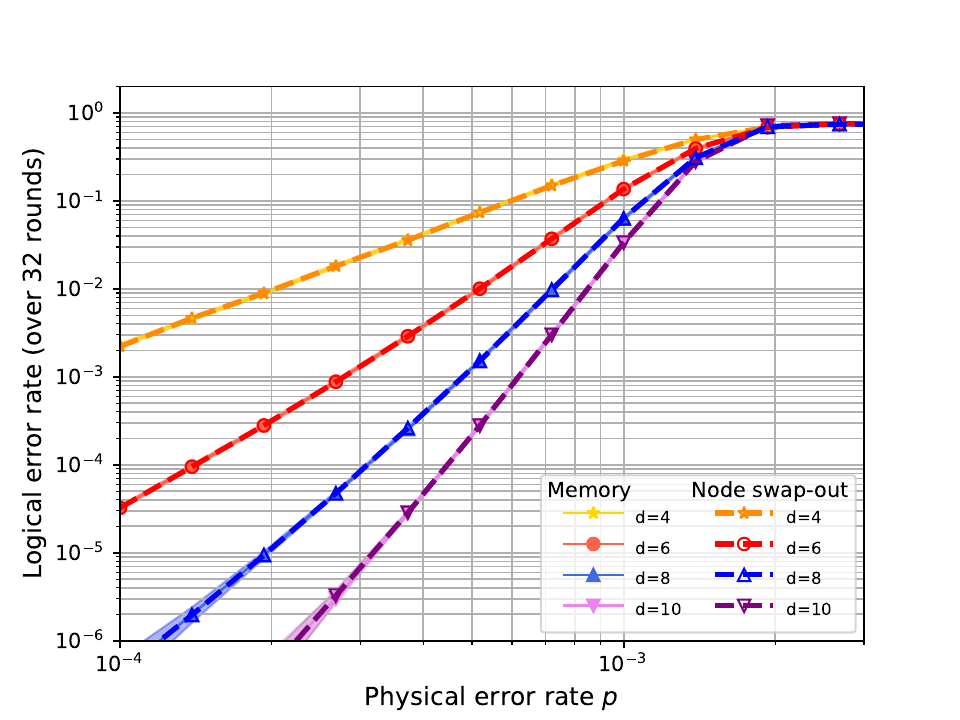}
    \caption{}
    \label{fig:x_a}
  \end{subfigure}
  \hfill
  \begin{subfigure}[b]{0.95\linewidth}
    \includegraphics[width=\textwidth]{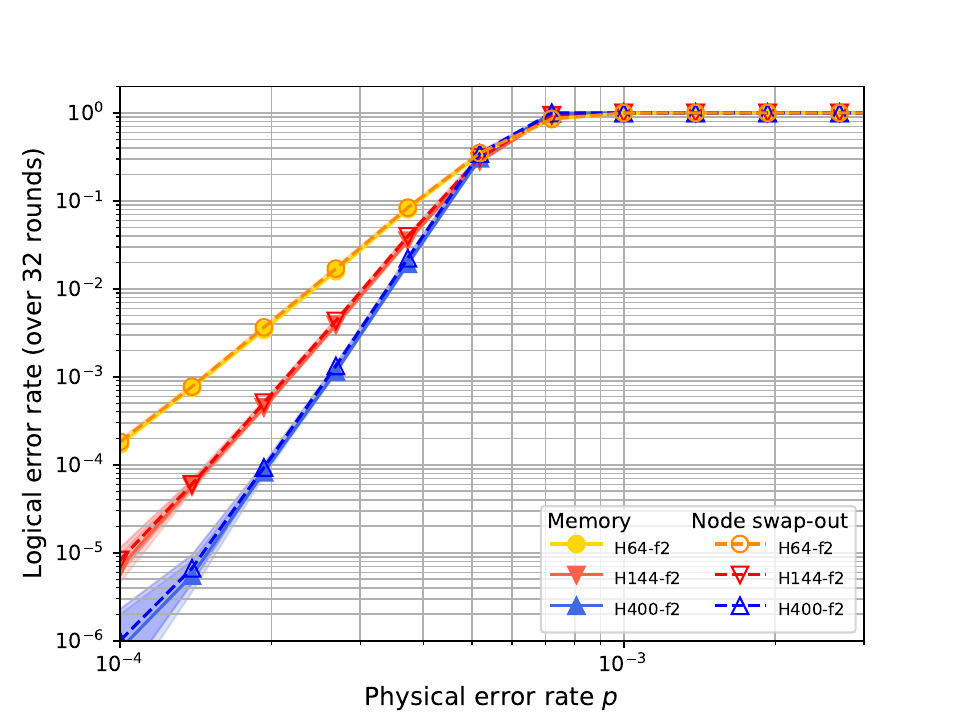}
    \caption{}
    \label{fig:x_b}
  \end{subfigure}
  \caption{(a) Logical error rate of toric codes of distance $d \in [4,6,8]$, with nodes of $n_q=16$ qubits. (b) The same results but for distributed semi-hyperbolic ($f=2$) Floquet codes. The swap-out procedure (dashed) has a minimal impact on the logical error rate compared to the quantum memory experiment (solid).
  The logical error rates were calculated for all $k$ logical observables after $r=32$ code rounds.}
  \label{fig:x}
\end{figure}

We first analyse the impact of the proposed node swap-out procedure on the logical error rate $P_L$ of the toric and semi-hyperbolic Floquet codes. We compare the logical error rate of a swap-out operation to a quantum memory experiment, without swap-out. For both, the logical error rate of the QEC codes was calculated after $r=32$ code rounds, and implemented in a network of QPUs, each of $n_q=16$ qubits. To test the swap-out procedure, a swap-out is performed after the $16$th code round. The swap-out was performed to replace the largest node in the network, by number of data qubits, as this was expected to have the worst performance. 
Figure~\ref{fig:x_a} shows the logical error rate of distributed toric codes with (solid) and without (dashed) performing swap-out. The results show there is a minimal impact on both the threshold and logical error rates. Similarly, Fig.~\ref{fig:x_b} shows the same simulation for distributed semi-hyperbolic Floquet codes, with similar results. As the qubit teleportation performed during swap-out is applied to each of the data qubits transversally, physical errors that occur during swap-out can not spread and hence have no impact on code distance. Combined with the fact that node swap-out is expected to be an infrequent operation, we see that scheduled node swap-out has a minimal impact on logical error rates.

\subsection{Analysing unscheduled node failure} \label{sec:drop_out1}
We analyse the performance of the distributed QEC codes under unscheduled node failure. Firstly, we show that node failure is catastrophic for monolithic devices, but can be reduced for distributed systems. Then, we identify regimes for which node failure events are the dominant cause of logical errors, instead of circuit noise. However, reducing QPU sizes or increasing the code distance can reduce the impact of node failure events. In Section~\ref{sec:drop_out2} we then show that with the correct choice of distributed implementation and quantum code, the impact of node failure on logical error rates can be suppressed to a negligible level.

\begin{figure}[h!tbp]
    \centering
    \includegraphics[width=1\linewidth]{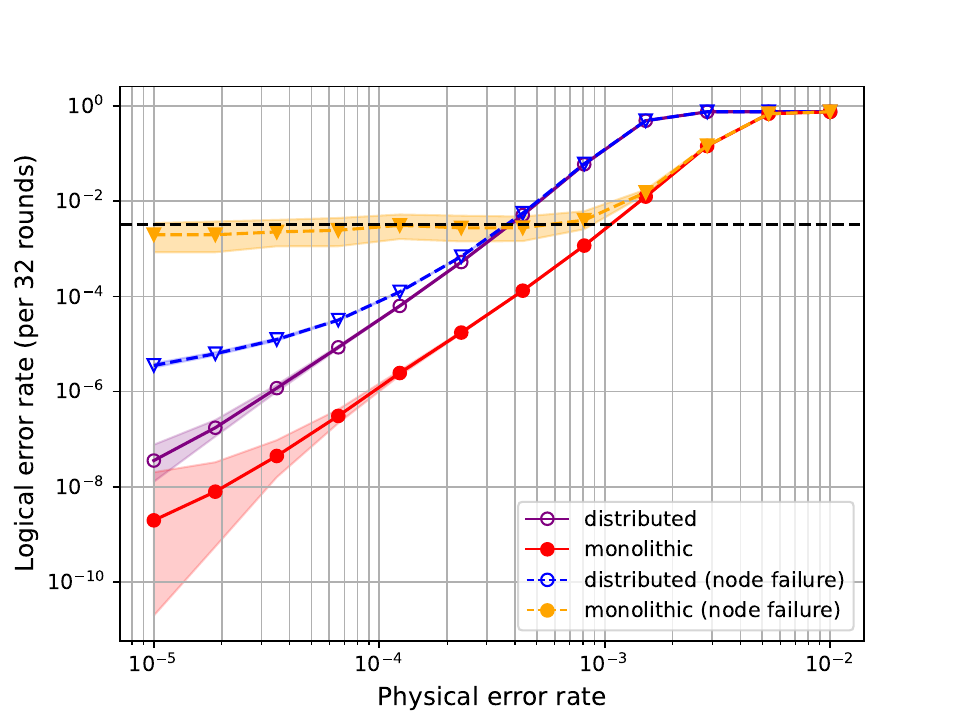}
    \caption{Logical error rate for the unrotated toric code ($d=6$). Results are shown for a monolithic system (green and yellow) and for the code distributed over a set of $n_q=16$ qubit devices (blue and red). Results under node failure rates of $p_{\textrm{dropout}}=1\times10^{-4}$ show that the distributed QEC code maintains error correction capabilities, below the analytical floor of a monolithic device (dashed black).}
    \label{fig:new_plot}
\end{figure}

While node failures are expected to be a rare event, occurring with a low probability, such errors affect a large number of qubits simultaneously, increasing their overall impact. Therefore, catastrophic node failure in monolithic systems can set a floor on the achievable logical error rate. Figure~\ref{fig:new_plot} shows the performance of a $d=6$ toric code, with its logical error rate calculated on a monolithic (yellow) and distributed (red) system without node failure.

We observe that the distributed QEC code starts correcting logical errors at a lower value of $p$, compared to the QEC code implemented on a monolithic system, due to the presence of elevated noise from non-local operations, as well as additional idling from the distributed compilation. However, both approaches appear to have similar error suppression scaling performance, up to a fixed shift in pseudothreshold. We note, however, that a distributed system may be more scalable in construction, as well as potentially allowing for the efficient implementation of the long-range connectivity required for high-rate codes \cite{sutcliffe2025distributed}.

Both distributed and monolithic systems were also simulated with a fixed per-round probability of node failure $p_{\textrm{dropout}}=1\times10^{-4}$. 
When modelling systems with node dropout, the monolithic approach achieves significantly worse performance than the distributed approach, with the logical error rate tailing off at $P_L\approx 3 \times 10^{-3}$. This limit can be described by the probability of at least one node failure event over any of the $r$ rounds for which error correction was performed:
\begin{equation} 
   P_L(p_\text{{dropout}},r) =   1-(1-p_\text{{dropout}})^{r} \label{eq:2}
\end{equation}
This analytical value is shown as a dashed black line in Fig.~\ref{fig:new_plot} and represents a floor in the logical error rate for monolithic systems.\footnote{Specifically, Eq.~\ref{eq:2} represents the probability of a catastrophic node failure. For an exact floor on the logical error rate of $k$ logical qubits, should be multiplied by a factor $1-1/2^k$, to account for the randomised state of each of the $k$ logical qubits.} In contrast, the distributed system under node failure (blue) was able to achieve a logical error rate below the analytical floor for monolithic systems. At $p = 1 \times 10^{-4}$, the logical error rate is found to be around $100$ times lower than the monolithic system, but also approximately $\times5$ larger than a distributed system without node failure. This suggested that the distributed approach is able to correct node failure events, but that they are still a significant source of logical errors. The results suggest that distributing a quantum code over a set of intermediate sized quantum devices is a valid approach to overcome node failure events. 

For a distributed quantum code, node failure will cause correlated errors on a subset of qubits. However, as long as node failures do not cause a sufficiently large number of physical qubit errors, such errors can be corrected using standard decoding techniques \cite{higgott2022pymatching,roffe2020bposd}. To assess the impact of node failure events on logical error rates, we consider noise models in which node failure ($p_\text{{dropout}}$) and circuit-noise ($p_\text{{circ}}$) are considered in isolation. The three models considered were: \textit{circuit noise} only ($p_\text{{circ}} = p, p_\text{{dropout}}=0$), \textit{node failure} only ($p_\text{{circ}} = 0, p_\text{{dropout}}=p/100$), and \textit{combined} noise model with both applied simultaneously ($p_\text{{circ}} = p, p_\text{{dropout}}=p/100$). The circuit noise model includes both local circuit noise $p_l = p$, as well as non-local noise during Bell pair generation $p_{nl} = 10p$.
The node dropout probability of $p/100$ was applied per code round. We use this to model node failure as a rare event, with a probability that is small in comparison to the physical error rate. 

\begin{figure}[h!tbp]
    \centering
    \includegraphics[width=1\linewidth]{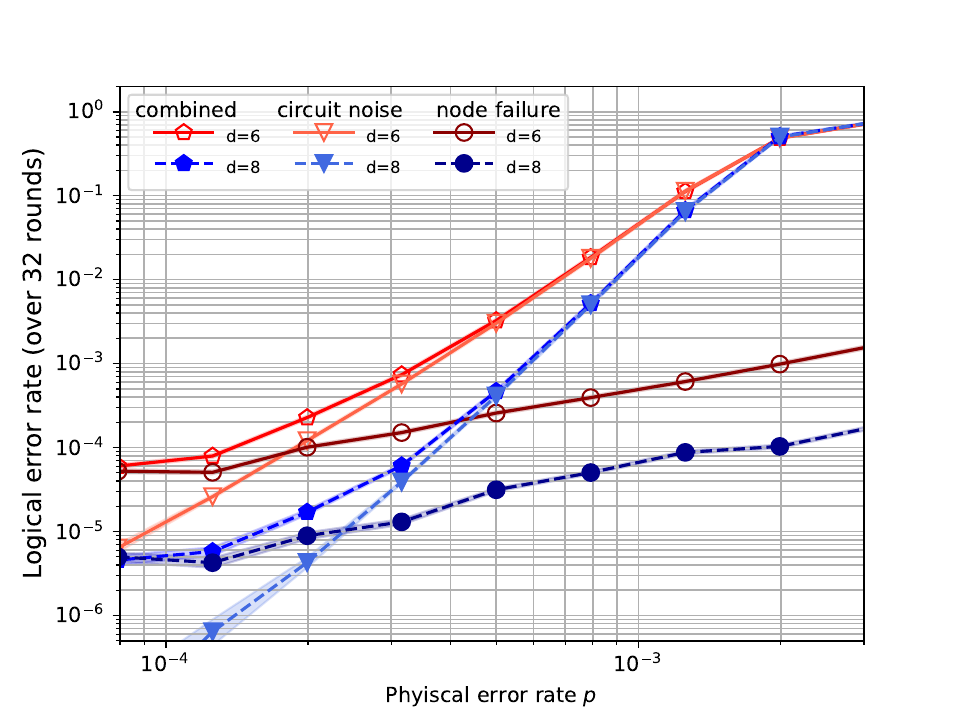}
    \caption{Logical error rate of the toric code ($d=6,8$) distributed over a $n_q=48$ qubit device.} 
    \label{fig:res4}
\end{figure}

Figure~\ref{fig:res4} shows the logical error rate of distributed toric codes ($d=6,8$), implemented over devices of $n_q = 48$ qubits. For this implementation, the $d = 6$ and $d=8$ toric codes are distributed over only four and eight QPUs, respectively. The results show that node failure and circuit noise both have separate scaling behaviour in terms of logical error rate, and that for the examples considered, a crossover regime exists between the sources of logical error. For the \textit{combined} model, this crossover represents the value of $p$ for which node failure prevents exponential logical error suppression below the threshold. For example, the distance $d=6$ toric code, the error rates for circuit noise and node failure errors crossover at $p = 1.5 \times 10^{-4}$. 
Therefore, in quantum devices prone to both node failure and circuit noise, the achievable logical error rate will be limited by the maximum of these two sources of error. This crossover occurs despite node dropout probability remaining a constant fraction ($p/100$) of circuit-level noise. We suggest that under node failure, a finite fraction of device failures can be tolerated for a given quantum code, as long as the number of qubits lost is small in comparison to the size of the quantum code.
To mitigate the impact of these events, we argue that increasing the code distance (and hence number of QPUs) lowers the logical error rate for both failure types. Further, decreasing the size of the QPU (not shown) also reduces the logical error rate under unscheduled node failure. This means that under the correct selection of the quantum code and device size, logical errors due to node failure can be prevented.

\subsection{Performance under unscheduled node failure} \label{sec:drop_out2}

\begin{figure}[h!tbp]
    \centering
    \includegraphics[width=1\linewidth]{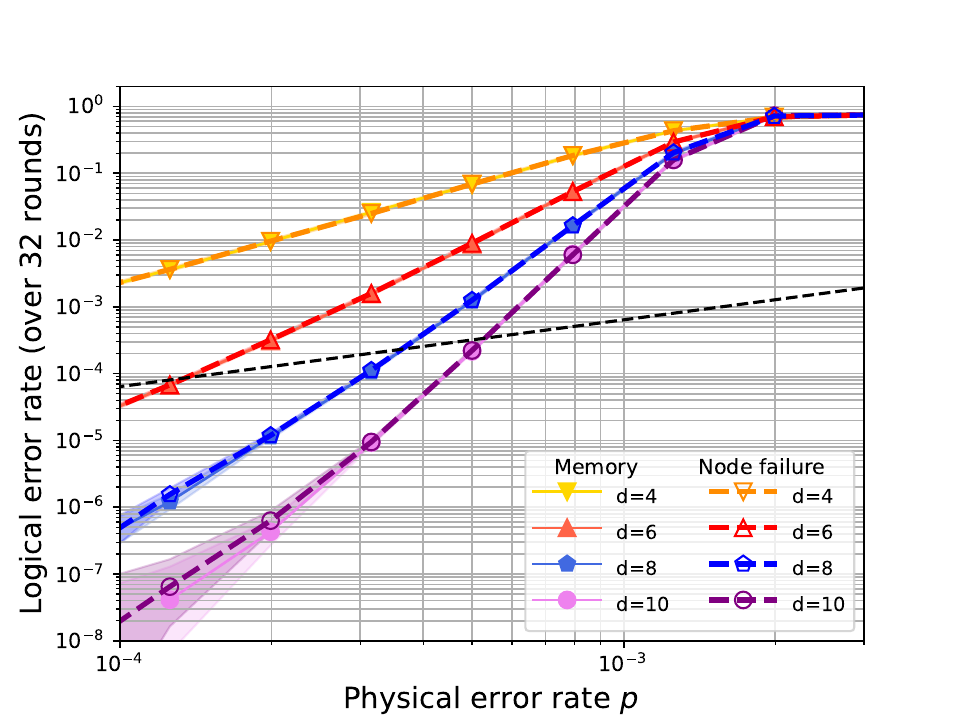}
    \includegraphics[width=1\linewidth]{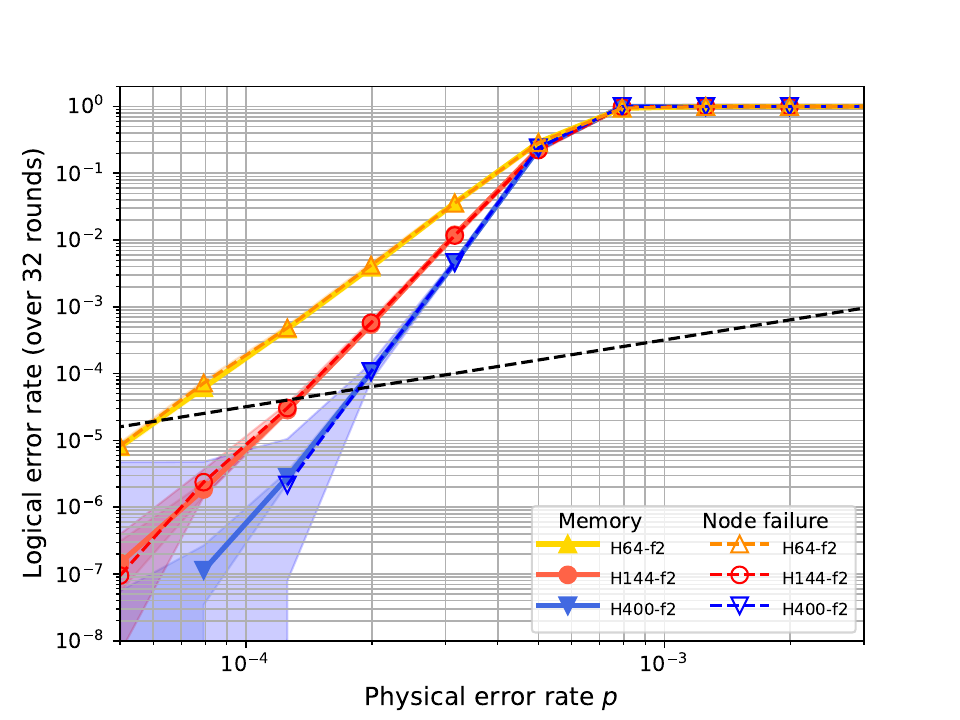}

    \caption{The impact of physical error rate $p$ and node failure $p/100$ on logical error rates for (a), unrotated toric codes with $d \in [6,\dots,12]$ and (b), semi-hyperbolic Floquet codes. Also shown is the analytical floor on logical error rate (black dashed line) for a single monolithic device with node failure $p/100$ per round.}
    \label{fig:res2}
\end{figure}

By assessing the logical error rate from only node failure, we find that for a certain size of QPU, logical errors due to node failure are suppressed to a higher degree for quantum codes of increasing code distance. This is because larger QEC codes are implemented over more nodes, increasing their tolerance to specific node failures. Fig.~\ref{fig:res2} shows the achieved logical error rate against the physical error rate $p$ of the distributed toric code (Fig.~\ref{fig:res2}a) and distributed hyperbolic Floquet code (Fig.~\ref{fig:res2}b). These codes are implemented over devices of $n_q=16$ qubits, and for the selected node size and partitioning, we find that there is a minimal impact on the logical error rate from unscheduled node failure.  

Also shown (black dashed line) is the probability described in  Eq.~\ref{eq:2} for varied $p_{\text{dropout}}$ and $r=32$ code rounds. For a $d=10$ toric code, we would therefore expect a distributed QEC code to outperform a monolithic device at a physical error rate of $0.05\%$.
A key benefit of the proposed approach is that increasing the size of the QEC code is efficient for suppressing both logical errors due to node failure and physical qubit errors. This is in contrast to approaches that require code concatenation to encode the logical information held at each QPU in a second level of erasure-correcting codes \cite{xu2022distributed}. Such approaches require a qubit overhead of at least $\times 6$ to tolerate node failure rate. 
Further, the choice of base code is limited by the capabilities and size of the QPUs. In the proposed distributed QEC approach, node failures are corrected using standard QEC techniques, allowing for the design of a fault-tolerant quantum computer to have flexibility in the choice of quantum code. Further, the total qubit overhead is given by $\lceil\sqrt{n_q}\rceil$ qubits per QPU, which are required for entanglement generation by the QPIs. However, this overhead is significantly less than $\times 6 $ even for the trivial case of QPUs of $n_q=1$. This is valid as long as the code distance and QPU size are selected such that node failure errors do not dominate logical error rates. 

\section{Conclusion and future work}

We have shown that distributed quantum error correcting codes implemented over a modular quantum network can tolerate node failure and dynamic node replacement. Firstly, we have shown how nodes can be swapped out and replaced with minimal impact on the logical error rate. We further show that distributed QEC codes are robust to catastrophic node failure using standard error correction techniques, unlike standard monolithic implementations. The proposed scheme is also significantly more efficient than previously proposed approaches to mitigating node failure, which incur a $\times6$-$9$ qubit overhead. For a per-round node failure probability of $p/100$, we show that distributed toric and Floquet codes will outperform monolithic implementations from physical error rates of $10^{-4}$-$10^{-3}$, representing physically relevant physical error rates.

One area of further work to consider the network implementation. In the future, it will be important to understand how to minimise the additional network connectivity required to support dynamic node substitution. Another approach to consider may be operating the quantum code with an additional boundary, to avoid replacing nodes and rather operate the code with a subset of qubits removed.
Additionally, we assume node failure is identified through sideband information channels, which inform a control system that node replacement is required. Whilst for certain platforms and error processes this is possible \cite{coble2025correctionchainlossestrapped,binney2026distinguishing}, directly observing node failure from syndrome measurements through the aid of a classifier would support a more universal approach.

\section*{Acknowledgement}
We thank Bhargavi Jonnadula, Alexandra E. Moylett, and Ed Wood for providing us with helpful comments during the preparation of this manuscript. We also thank our colleagues at Nu Quantum for helpful discussions, and thank \hbox{Carmen Palacios-Berraquero} for providing an environment where this work was possible.

\bibliographystyle{IEEEtran}
\bibliography{IEEEabrv,bib}

\clearpage

\end{document}